# FIRST INDIRECT X-RAY IMAGING TESTS WITH AN 88-mm DIAMETER SINGLE CRYSTAL

A.H. Lumpkin[1] and A.T. Macrander[2]

[1]Fermi National Accelerator Laboratory, Batavia, IL 60510 USA
[2]Argonne National Laboratory, Argonne, IL 60439 USA

## ABSTRACT

Using the 1-BM-C beamline at the Advanced Photon Source (APS), we have performed the initial indirect x-ray imaging point-spread-function (PSF) test of a unique 88-mm diameter YAG:Ce single crystal of only 100-micron thickness. The crystal was bonded to a fiber optic plate (FOP) for mechanical support and to allow the option for FO coupling to a large format camera. This configuration resolution was compared to that of self-supported 25-mm diameter crystals, with and without an Al reflective coating. An upstream monochromator was used to select 17-keV x-rays from the broadband APS bending magnet source of synchrotron radiation. The upstream, adjustable Mo collimators were then used to provide a series of x-ray source transverse sizes from 200 microns down to about 15-20 microns (FWHM) at the crystal surface. The emitted scintillator radiation was in this case lens coupled to the ANDOR Neo sCMOS camera, and the indirect x-ray images were processed offline by a MATLAB-based image processing program. Based on single Gaussian peak fits to the x-ray image projected profiles, we observed a 10.5 micron PSF. This sample thus exhibited superior spatial resolution to standard P43 polycrystalline phosphors of the same thickness which would have about a 100-micron PSF. This single crystal resolution combined with the 88-mm diameter makes it a candidate to support future x-ray diffraction or wafer topography experiments.

Key words: x-rays, PSF, single crystal scintillator

Index:

[1]Work at Fermilab partly supported by Fermi Research Alliance, LLC under Contract No. DE-AC02- 07CH11359 with the United States Department of Energy.

[2]Work at ANL supported by US DOE, Office of Science, under contract No. DE-AC02-06CH11357.

# INTRODUCTION

For a number of years the indirect x-ray imaging field has recognized the trade-off that must be made on converter screen spatial resolution and efficiency. This has been particularly true for polycrystalline phosphors such as $Gd_2O_2S$:Tb, or P43, where the resolution at full-width-at-half-maximum intensity (FWHM) is approximately equal to the screen thickness for 10-20 keV x-rays (Naday *et al.*,1994). Generally, a thinner screen is used when improved resolution is needed with the concomitant decrease in screen efficiency. If an Al reflective coating is added on the front surface for light collection efficiency, this will also impact the resolution. It has been observed (Tou *et al.*, 2007; Lumpkin *et al.*, 2011) that superior resolution can be obtained with single crystal scintillators of comparable thickness, although there still may be an optimal efficiency depending on the materials used. In this case we have chosen a set of rare earth garnet single crystals of 100-µm thickness, with and without the presence of a fiber optic plate (FOP). The crystal type is cerium-doped yttrium aluminum garnet, YAG:Ce, and one sample is a unique 88-mm diameter single crystal bonded to a 90-mm diameter FOP. We report the first point spread function (PSF) tests of this sample in a lens-coupled configuration using the x-ray beam in the APS 1-BM-C beamline. This scintillator crystal's combined high resolution and large diameter make it a candidate for x-ray crystal diffraction studies or wafer topography studies. It also has potential use in x-ray phase contrast imaging with a large format CCD system (Lumpkin *et al.*, 2016).

# EXPERIMENTAL TECHNIQUES

The APS is a hard x-ray user facility based on synchrotron radiation generated in a 7-GeV electron storage ring. The parameters of the APS bending magnet source of x-rays are given in Table 1. The normal stored electron beam current is 100 mA with a critical x-ray energy at 19.5 keV.

**TABLE 1.** Parameters of the APS Bending Magnet Source

| Parameter | Value |
|---|---|
| Storage Ring Electron Energy | 7.0 GeV |
| Storage Ring Electron Current | 100.0 mA |
| Bend Radius | 38.96 m |
| Peak Magnetic Field | 0.6 Tesla |
| Critical X-ray Energy | 19.5 keV |
| Horizontal Source Size | 198 µm FWHM |
| Vertical Source Size | 78 µm FWHM |
| Vertical Source Divergence at 11 keV | 157 µ rad FWHM |
| Flux at 19.5 keV | $1.12 \times 10^{13}$/sec/0.1%BW/mrad(H) |

TABLE 2. Components of the Beamline and distance from BM source.

| Component | Distance (meters) |
|---|---|
| Double Be window | 22.0 |
| white Beam Slits | 22.5 |
| Fast Shutter | 24.0 |
| Double Crystal Monochromator | 27.5 |
| Be Window | 30.5 |
| Granite Optical Bench | 35.5 |
| Be Window | 37 |
| Be Window | 53.5 |
| Huber Diffractometer | 55 |
| Optical Table | 57 |

The experiments were performed in station 1-BM-C beamline at the APS with the beam-line components listed in Table 2 (Macrander *et al.*, 2015), and these included the use of a double crystal monochromator to select the x-ray energy at 17 keV. The experiment was based at the 57-m point on an optical table. The 25-mm diameter crystals were mounted in Thorlabs lens holders and attached to a Thorlabs magnetic kinematic mount. The large crystal/FOP was mounted in a rotating holder with 98-mm diameter which had a hinged flip in/out feature. The experimental setup is shown in Fig. 1 with a small holder in place and the large holder flipped out of the path. The optical imaging system that viewed the crystal surface consisted of a 2.5x objective lens and an ANDOR Neo 5.5 Mpixel sCMOS camera with 6.5-µm square pixel sizes. A small piezo drive stage was used to set the z position of the crystal holder for optimum optical focus based on the sharpness of the collimator edge in the x-ray image.

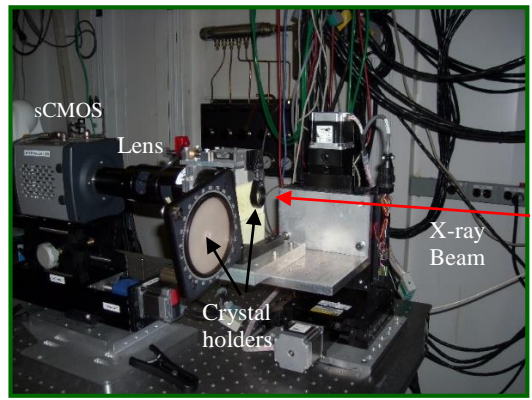

Figure 1: Photograph of the stages, crystal holders, and ANDOR camera configured on the 1-BM-C beamline optics table. A yellow alignment screen was temporarily in place.

The magnification resulted in a spatial calibration factor of 6.5/2.5= 2.60 µm/pixel based on the 1 mm by 1 mm collimated image size. The digital images were recorded and processed online with ImageJ. The images were also processed offline with a MATLAB-based program. The basic test involved the use of the upstream Mo collimator vertical aperture that was scanned from 1 mm to 20 µm while holding the horizontal size at 1 mm width. The vertical collimators were shown to have a 5-µm offset from zero, so we subsequently subtracted this value from the vertical readings. The PSF was estimated from the image FWHM (2.35 times the fit sigma value) after the quadrature subtraction of the collimator aperture size. The collimator images are flat topped, but at the key small aperture sizes a Gaussian profile was a good approximation to the projected profiles.

One of the main objectives was to obtain first indirect x-ray images from the 88-mm diameter crystal bonded to the 90-mm diameter FOP. This unique crystal was prepared by Crytur, LLC under a custom order from FNAL, and it is shown in comparison to the standard 25-mm diameter crystal and a quarter in Fig. 2. Another perspective is shown in Fig. 3 with the hand-held sample.

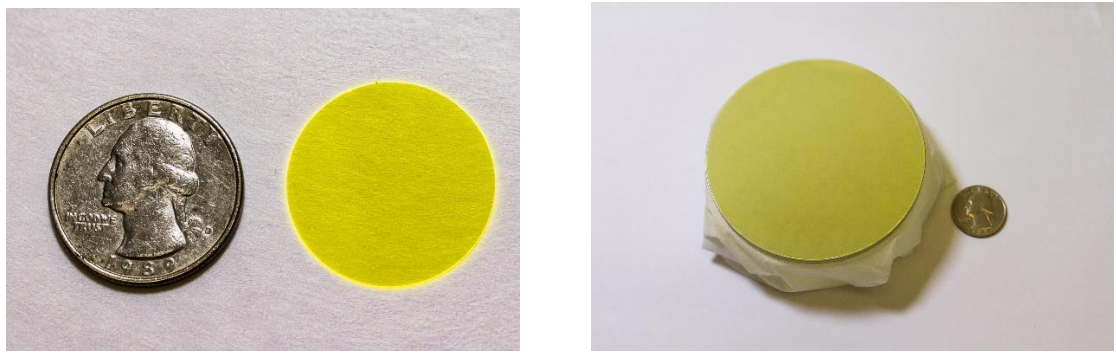

Figure 2: Comparsion of 25-mm diameter crystal to a quarter (L) and the 88-mm diameter crystal to the same quarter (R). The latter crystal is more than 12 times bigger in area. (Photos by E. McCrory-FNAL)

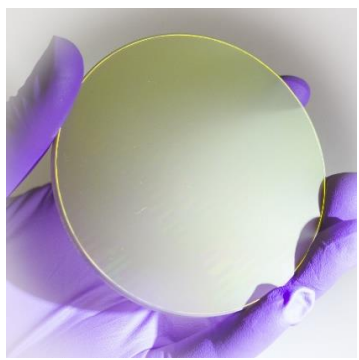

Figure 3: Photograph of the 88-mm diameter crystal on the FOP. (photo by E. McCrory).

## EXPERIMENTAL RESULTS

A series of collimator settings was used to generate slit images of varying height that were recorded by the indirect x-ray imaging system. Data were taken with the 100-µm thick crystals with and without the Al reflective coating to see if measurable resolution effects were present. Examples of the two images are shown in Fig. 4 for an adjusted aperture size of 20 µm. For the one crystal there was an apparent Ce dopant defect that is reflected in the projected image from the region of interest (ROI). This was addressed with a two-Gaussian fit that reproduces the data well. The main profile has a size of sigma =3.7 pixels, or 21.7 µm FWHM, but underestimated the peak. If we shifted the ROI to avoid the dopant defect, we obtained a single Gaussian fit result of 27.3 µm FWHM. The Al-coated crystal's image profile was fit to a single Gaussian since no defect was involved. The result was sigma = 4.6 pixels, or 27.0 µm FWHM. These latter two values indicate that with lens coupling the light signal is enhanced by the Al without a significant loss of resolution, unlike the polycrystalline cases where more scattering is involved. Subtracting out the collimator size in quadrature from the total observed image sizes, gives values of 18.5 µm FWHM.

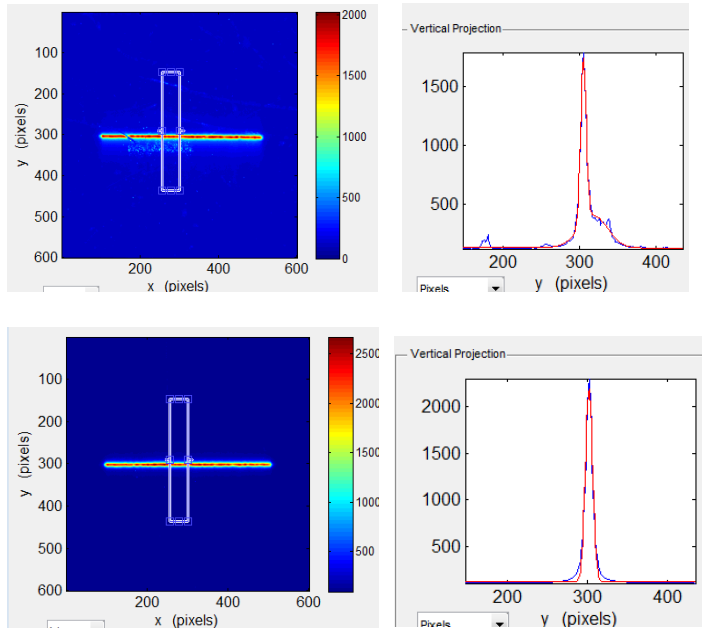

Figure 4: Examples of the slit images with the 100-µm YAG without Al coating (Top) and with coating (bottom). The vertical projected profiles are fit and shown at the right of the images.

An example image is shown in Fig. 5 for the 88-mm diameter crystal bonded to the FOP for the actual 20 µm vertical slit height which give an x-ray image vertical profile with sigma = 3.8 pixels, or FWHM size of 22.3 µm. If we subtract out the collimator FWHM size of 20 µm, the system PSF is estimated at 10.5 µm. In comparison, a typical P43 polycrystalline phosphor of 100-µm thickness would be expected to have a PSF of 100 µm. Such a term would dominate the system PSF in this case. In Fig. 4 and Fig. 5 we can compare collimator data of the Al coated and FOP bonded crystals. The FOP-bonded option has detectably better resolution in the lens coupled configuration in our tests after collimator size subtraction. This is also shown in Fig. 6 (a) for the smaller apertures.

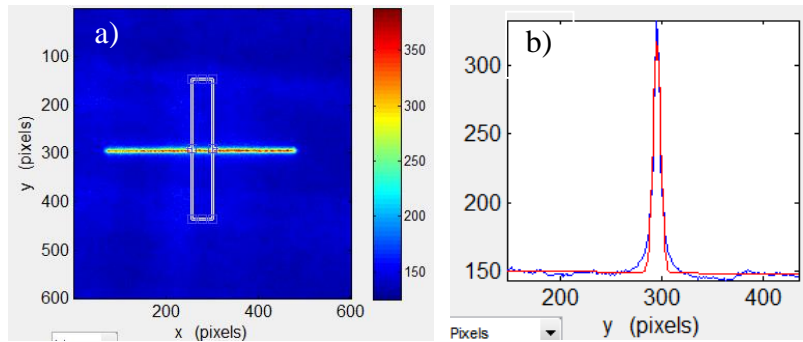

Figure 5: An example x-ray image (a) from the 88-mm diameter YAG:Ce crystal with the effective collimator slit vertical size of 20 µm and horizontal size of 1 mm. The projected vertical profile (blue) and fit profile (red) is shown in b).

In Figure 6 (b), we plot the variation in the observed slit image vertical profiles with the FOP coupled crystal as compared to the slit settings (blue). The limiting resolution appears to just be sensed at the last collimator setting shown.

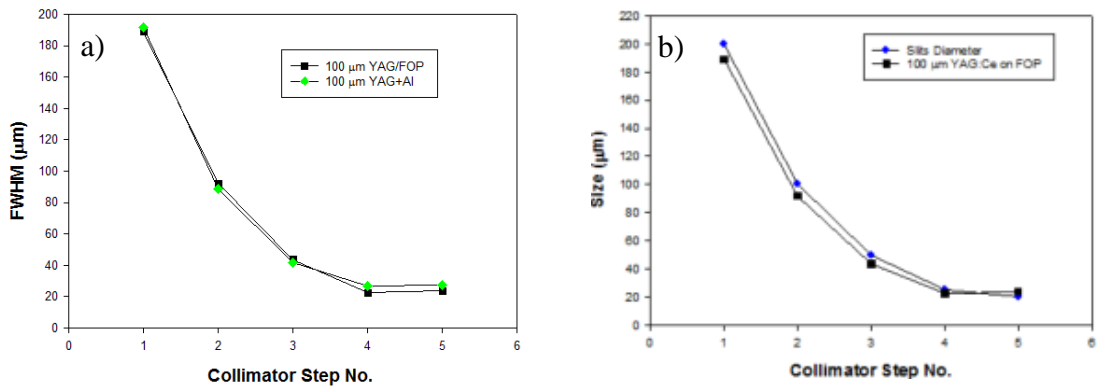

Figure 6: a) Comparison of the 100-µm thick YAG:Ce crystals, one with Al coating and one bonded to an FOP and b) Comparison of the slit width value and the YAG:Ce/FOP profile FWHM value.

In addition, we can perform a simple comparison in Fig. 7 of these lens-coupled results with data taken with fiber-optic-coupled single crystals in a large format camera [Lumpkin *et al*., 2016]. We use the term effective optical thickness to account for the Al coating as a reflector on the P43 and two of the crystals. This means some of the light is traversing a longer distance before the FOP, and thus the depth of focus of the FO bundle comes into play. The P43 phosphor data (red square) is the reference case with an effective optical thickness of ~70 µm and a PSF of about 82 µm. The single crystal data with similar coupling fall below this value on the black line whose slope is consistent with the blurring of the PSF values with larger effective optical thicknesses. The APS lens coupled data (green), fall noticeably below this curve. However, the fiber optic coupling is more efficient for signal collection so there is a trade-off still to be considered.

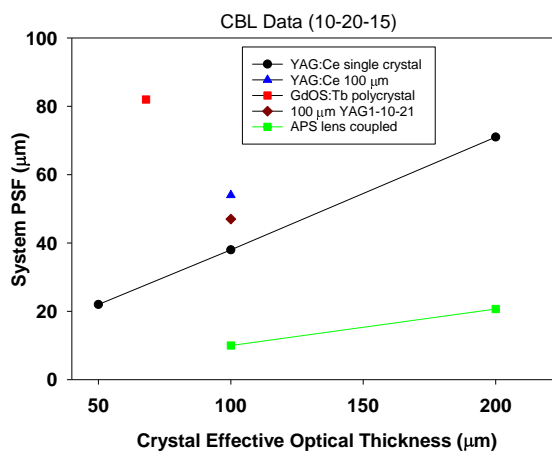

Figure 7: Comparison of the PSFs for P43 with FO coupling, single crystals with FO coupling, and single crystals using lens coupling at APS.

## SUMMARY

In summary, we have performed initial indirect x-ray imaging tests with a set of 100-µm thick single crystals with and without Al coating and with and without an FOP. With lens coupling the 88-mm diameter crystal provided a PSF 10 times better than that expected of a polycrystalline P43 screen with the same thickness (Lumpkin, 2016). The combined resolution and large area should make it applicable to crystal diffraction tests or wafer topography experiments. Applications in biomedical imaging are also being pursued.

## ACKNOWLEDGEMENTS

The first author acknowledges the support of E. Ramberg of PPD/FNAL and N. Eddy of AD/FNAL and the engineering support of J. Fitzgerald AD/FNAL. We also acknowledge the loan of a 25-mm diameter YAG:Ce crystal from W. Berg (ANL) for these tests. This is the first test of "Katherine's Krystal" with its 88-mm diameter.